\documentclass[acmtog]{acmart}

\usepackage{booktabs}
\usepackage{epsfig}
\usepackage{graphicx}
\usepackage{amsmath}
\usepackage{amssymb}
\usepackage{caption}
\usepackage{multirow}

\acmJournal{TOG}
\acmVolume{37}
\acmNumber{4}
\acmArticle{112}
\acmYear{2018}
\acmMonth{8}

\setcitestyle{square}
\newcommand{\markedit}[1]{{\color{red}#1}}
\renewcommand{\markedit}[1]{#1}

\newcommand{\afterfigure}{\vspace{-1em}}
\def\datasetname{\textsc{AVSpeech}}

\copyrightyear{2018} 
\acmYear{2018} 
\setcopyright{rightsretained} 
\acmConference[SIGGRAPH '18 Technical Paper]{Special Interest Group on Computer Graphics and Interactive Techniques Conference Technical Paper}{August 12--16, 2018}{Vancouver, BC, Canada}
\acmBooktitle{SIGGRAPH '18 Technical Paper: Special Interest Group on Computer Graphics and Interactive Techniques Conference Technical Paper, August 12--16, 2018, Vancouver, BC, Canada}
\acmDOI{10.1145/3197517.3201357}
\acmISBN{978-1-4503-5763-0/18/08}

\begin{document}
\title[Looking to Listen at the Cocktail Party]{Looking to Listen at the Cocktail Party: \\ A Speaker-Independent Audio-Visual Model for Speech Separation}

\author{Ariel Ephrat} \affiliation{Google Research} \affiliation{The Hebrew University of Jerusalem, Israel}
\author{Inbar Mosseri} \affiliation{Google Research}
\author{Oran Lang} \affiliation{Google Research}
\author{Tali Dekel} \affiliation{Google Research}
\author{Kevin Wilson} \affiliation{Google Research}
\author{Avinatan Hassidim} \affiliation{Google Research}
\author{William T. Freeman} \affiliation{Google Research}
\author{Michael Rubinstein} \affiliation{Google Research}
\renewcommand\shortauthors{Ephrat, A. et al}
\authorsaddresses{}

\begin{abstract}
We present a joint audio-visual model for isolating a single speech signal from a mixture of sounds such as other speakers and background noise. Solving this task using only audio as input is extremely challenging and does not provide an association of the separated speech signals with speakers in the video. In this paper, we present a deep network-based model that incorporates both visual and auditory signals to solve this task. The visual features are used to ``focus'' the audio on desired speakers in a scene and to improve the speech separation quality. To train our joint audio-visual model, we introduce \datasetname, a new dataset comprised of thousands of hours of video segments from the Web. We demonstrate the applicability of our method to classic speech separation tasks, as well as real-world scenarios involving heated interviews, noisy bars, and screaming children, only requiring the user to specify the face of the person in the video whose speech they want to isolate. Our method shows clear advantage over state-of-the-art audio-only speech separation in cases of mixed speech. In addition, our model, which is speaker-independent (trained once, applicable to any speaker), produces better results than recent audio-visual speech separation methods that are speaker-dependent (require training a separate model for each speaker of interest).

\let\thefootnote\relax\footnotetext{The first author performed this work as an intern at Google.}

\end{abstract}
\begin{teaserfigure}
\centering
\includegraphics[width=\linewidth]{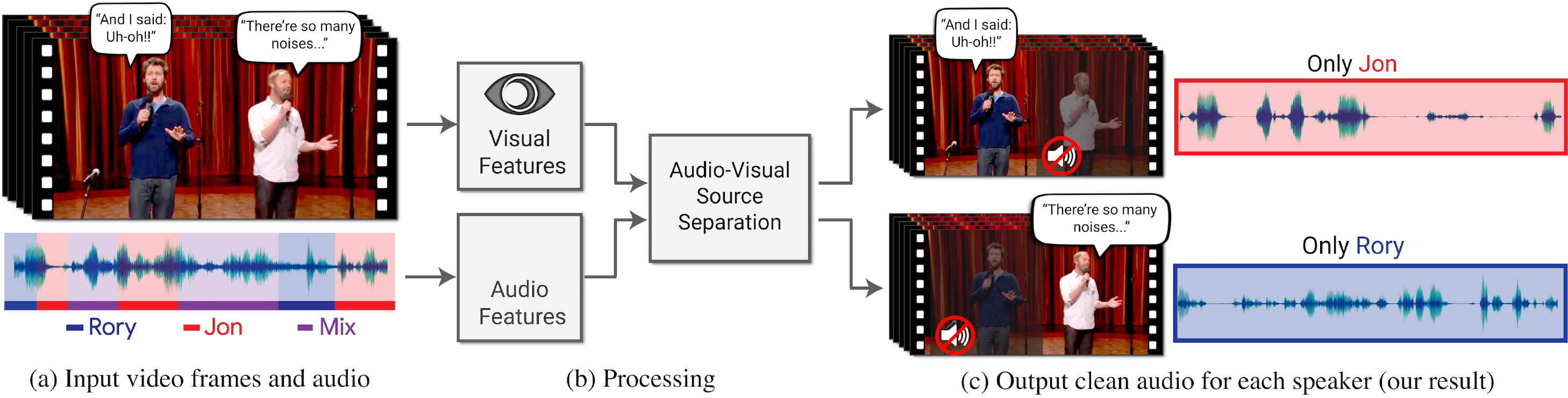}
\vspace{-.2in}
\caption{We present a model for isolating and enhancing the speech of desired speakers in a video. (a) The input is a video (frames + audio track) with one or more people speaking, where the speech of interest is interfered by other speakers and/or background noise. (b) Both audio and visual features are extracted and fed into a joint audio-visual speech separation model. The output is a decomposition of the input audio track into clean speech tracks, one for each person detected in the video (c). This allows us to then compose videos where speech of specific people is enhanced while all other sound is suppressed. Our model was trained using thousands of hours of video segments from our new dataset, \datasetname. The \emph{``Stand-Up''} video (a) is courtesy of Team Coco.\vspace{.1in}}
\label{fig:teaser}
\end{teaserfigure}


\maketitle

\section{Introduction} \label{sec:intro}

Humans are remarkably capable of focusing their auditory attention on a single sound source within a noisy environment, while de-emphasizing (``muting'') all other voices and sounds. The way neural systems achieve this feat, which is known as the \emph{cocktail party effect} \cite{cherry1953some}, remains unclear. However, research has shown that viewing a speaker's face enhances a person's capacity to resolve perceptual ambiguity in a noisy environment \markedit{\cite{Ma2009LipReadingAW,Golumbic2013VisualIE}}. In this paper we achieve this ability computationally.

Automatic speech separation---separating an input audio signal into its individual speech sources---is well-studied in the audio processing literature. Since this problem is inherently ill-posed, it requires prior knowledge or special microphone configurations in order to obtain a reasonable solution \cite{mcdermott2009cocktail}. In addition, a fundamental problem with audio-only speech separation is the \emph{label permutation problem}~\markedit{\cite{Hershey2016DeepCD}}: there is no easy way to associate each separated audio source with its corresponding speaker in the video~\cite{Yu2017PermutationIT, Hershey2016DeepCD}.

In this work, we present a joint audio-visual method for ``focusing'' audio on a desired speaker in a video. The input video can then be recomposed such that the audio corresponding to specific people is enhanced while all other sound is suppressed (Fig.~\ref{fig:teaser}). More specifically, we design and train a neural network-based model that takes the recorded sound mixture, along with tight crops of detected faces in each frame in the video as input, and splits the mixture into separate audio streams for each detected speaker. The model uses visual information both as a means to improve the source separation quality (compared to audio-only results), as well as to associate the separated speech tracks with visible speakers in the video. All that is required from the user is to specify which faces of the people in the video they want to hear the speech from.

\begin{figure*}
   \centering
   \includegraphics[width=\linewidth]{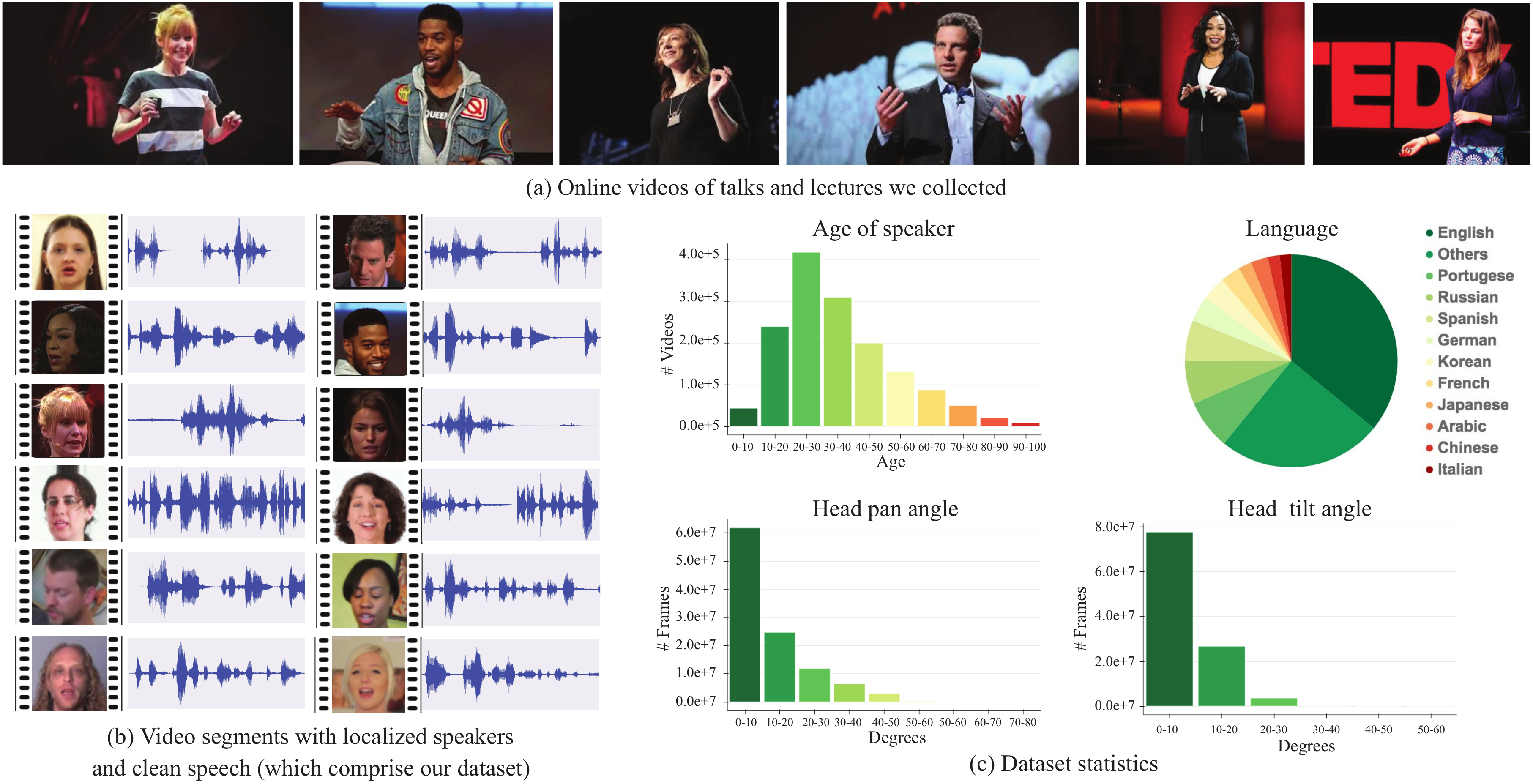}
   \vspace{-.2in}
   \caption{{\bf \textsc{\datasetname} dataset:} We first gathered a large collection of 290,000 high-quality, online public videos of talks and lectures (a). From these videos we extracted segments with clean speech (e.g. no mixed music, audience sounds or other speakers), and with the speaker visible in the frame (see Section \ref{sec:dataset} and Figure \ref{fig:dataset_pipeline}, for details of the processing). This resulted in 4700 hours of video clips, each of a single person talking with no background interference (b). This data spans a wide variety of people, languages, and face poses, with distributions shown in (c) (age and head angles estimated with automatic classifiers; language based on YouTube metadata). For a detailed list of video sources in our dataset please refer to the project web page.}
\label{fig:dataset}
\afterfigure
\end{figure*}

To train our model, we collected 290,000 high-quality lectures, TED talks and how-to videos from YouTube, then automatically extracted from these videos roughly 4700 hours of video clips with visible speakers and clean speech with no interfering sounds (Fig.~\ref{fig:dataset}). We call our new dataset \textsc{\datasetname}. With this dataset in hand, we then generated a training set of \emph{``synthetic cocktail parties''}---mixtures of face videos with clean speech, and other speech audio tracks and background noise.

We demonstrate the benefits of our approach over recent speech separation methods in two ways. First, we show superior results compared  to a state-of-the-art audio-only method on pure speech mixtures. Second, we demonstrate our model's capability of producing enhanced sound streams from mixtures containing both overlapping speech and background noise in real-world scenarios.

To summarize, our paper makes two main contributions: (a) An audio-visual speech separation model that outperforms audio-only and audio-visual models on classic speech separation tasks, and is applicable in challenging, natural scenes. To our knowledge, our paper is the first to propose a speaker-independent audio-visual model for speech separation. (b) A new, large-scale audio-visual dataset, \textsc{\datasetname},  carefully collected and processed, comprised of video segments where the audible sound belongs to a single person, visible in the video, and no audio background interference. This dataset allows us to achieve state-of-the-art results on speech separation and may be useful for the research community for further studies. Our dataset, input and output videos, and additional supplementary materials are all available on the project web page:
\url{http://looking-to-listen.github.io/}.

\section{Related work}
\label{sec:related}
We briefly review related work in the areas of speech separation and audio-visual signal processing.

\paragraph{Speech separation.}

Speech separation is one of the fundamental problems in audio processing and has been the subject of extensive study over the last decades. \citet{ss_overview} give a comprehensive overview of recent audio-only methods based on deep learning that tackle both speech denoising \cite{Weninger2015SpeechEW,Erdogan2015PhasesensitiveAR} and speech separation tasks.

Two recent works have emerged which solve the aforementioned label permutation problem to perform speaker-independent, multi-speaker separation in the single-channel case. \citet{Hershey2016DeepCD} propose a method called \emph{deep clustering} in which discriminatively-trained speech embeddings are used to cluster and separate the different sources. \citet{Hershey2016DeepCD} also introduced the idea of a permutation-free or \emph{permutation invariant} loss function, but they did not find that it worked well.  \citet{isik2016single} and \citet{Yu2017PermutationIT} subsequently introduced methods which successfully use a permutation invariant loss function to train a DNN.

The advantages of our approach over such audio-only methods are threefold: First, we show that the separation results of our audio-visual model are of higher quality than those of a state-of-the-art-inspired audio-only model. Second, our approach performs well in the setting of multiple speakers mixed with background noise, which, to our knowledge, no audio-only method has satisfactorily solved. Third, we jointly solve two speech processing problems: speech separation, and assignment of a speech signal to its corresponding face, which, thus far, have been tackled separately \cite{Hoover2017PuttingAF,hu2015deep,monaci2011towards}.

%
%
%
%
%
%
%
%
%

\paragraph{Vision and speech.}

There is increased interest in using neural networks for multi-modal fusion of auditory and visual signals to solve various speech-related problems. These include audio-visual speech recognition \cite{Ngiam2011MultimodalDL,mroueh2015deep,feng2017audio},  predicting speech or text from silent video (lipreading) \cite{ephrat2017improved,Chung2016LipRS}, and unsupervised learning of language from visual and speech signals \cite{Harwath2016UnsupervisedLO}. These methods leverage natural synchrony between simultaneously recorded visual and auditory signals.

Audio-visual (AV) methods have also been used for speech separation and enhancement \cite{hershey2001audio,hershey2004audio,Rivet2014AudiovisualSS,khan2016audio}. \citet{casanovas2010blind} perform AV source separation using sparse representations, which is limited due to dependence on active-alone regions to learn source characteristics, and the assumption that all the audio sources are seen on-screen. Recent methods have used neural networks to perform the task. \citet{Hou2017AudioVisualSE} propose a multi-task CNN-based model which outputs a denoised speech spectrogram as well a reconstruction of the input mouth region. \citet{gabbay2017visual} train a speech enhancement model on videos where other speech samples of the target speaker are used as background noise, in a scheme they call \emph{``noise-invariant training''}. In concurrent work, \citet{gabbay2018seeing} use a video-to-sound synthesis method to filter noisy audio.

The main limitation of these AV speech separation approaches is that they are \emph{speaker-dependent}, meaning a dedicated model must be trained for each speaker separately. While these works make specific design choices that limit their applicability only to the speaker-dependent case, we speculate that the main reason a speaker-independent AV model hasn't been pursued widely so far is the lack of a sufficiently large and diverse dataset for training such models --- a dataset like the one we construct and provide in this work. To the best of our knowledge, our paper is the first to address the problem of \emph{speaker-independent} AV speech separation. Our model is capable of separating and enhancing speakers it has never seen before, speaking in languages that were not part of the training set. In addition, our work is unique in that we show high quality speech separation on real world examples, in settings that previous audio-only and audio-visual speech separation work did not address.

A number of independent and concurrent works have recently emerged which address the problem of audio-visual sound source separation using deep neural networks. \cite{owens2018audio} train a network to predict whether audio and visual streams are temporally aligned. Learned features extracted from this self-supervised model are then used to condition an on/off screen speakers source separation model. \citet{Afouras18} perform speech enhancement by using a network to predict both magnitude and phase of denoised speech spectrograms. \citet{zhao2018sound} and \citet{gao2018learning} addressed the closely related problem of separating the sound of multiple on-screen objects (e.g. musical instruments).

\paragraph{Audio-visual datasets.}
Most existing AV datasets comprise videos with only a small number of subjects, speaking words from a limited vocabulary. For example, the CUAVE dataset \cite{Patterson2002MovingTalkerSF} contains 36 subjects saying each digit from $0$ to $9$ five times each, with a total of 180 examples per digit. Another example is the Mandarin sentences dataset, introduced by \citet{Hou2017AudioVisualSE}, which contains video recordings of 320 utterances of Mandarin sentences spoken by a native speaker. Each sentence contains 10 Chinese characters with equally distributed phonemes. The TCD-TIMIT dataset \cite{harte2015tcd} consists of 60 volunteer speakers with around 200 videos each. The speakers recite various sentences from the TIMIT dataset \cite{timit}, and are recorded using both front-facing and $30$-degree cameras. We evaluate our results on these three datasets in order to compare to previous work.

Recently, the large-scale Lip Reading Sentences (LRS) dataset was introduced by \citet{Chung2016LipRS}, which includes both a wide variety of speakers and words from a larger vocabulary. However, not only is that dataset not publicly available, but the speech in LRS videos is not guaranteed to be clean, which is crucial for training a model for speech separation and enhancement.

\section{\textsc{\datasetname} Dataset} \label{sec:dataset}

We introduce a new, large-scale audio-visual dataset comprising speech clips with no interfering background signals. The segments are of varying length, between 3 and 10 seconds long, and in each clip the only visible face in the video and audible sound in the soundtrack belong to a single speaking person. In total, the dataset contains roughly 4700 hours of video segments with approximately 150,000 distinct speakers, spanning a wide variety of people, languages and face poses. Representative frames, audio waveforms and some dataset statistics are shown in Figure \ref{fig:dataset}.
%
%
%
%

We collected the dataset automatically, since for assembling a corpus of this magnitude it was important not to rely on substantial human feedback. Our dataset creation pipeline collected clips from roughly 290,000 YouTube videos of lectures (e.g. TED talks) and how-to videos. For such channels, most of the videos comprise a single speaker, and both the video and audio are generally of high quality.

\begin{figure}[t]
   \centering
   \includegraphics[width=\linewidth]{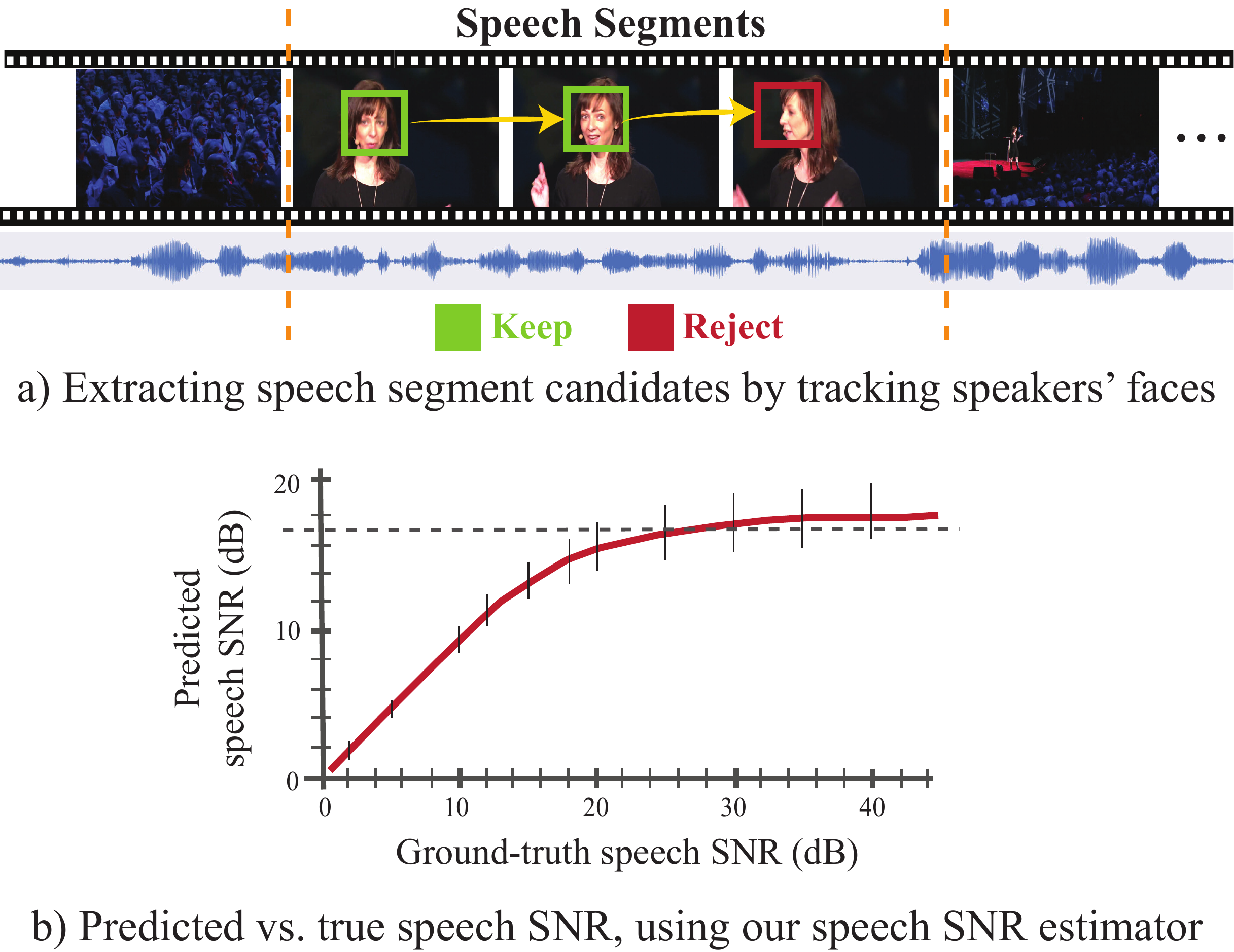}
   \vspace{-.2in}
   \caption{{\bf Video and audio processing for dataset creation:} (a) We use face detection and tracking to extract speech segment candidates from videos and reject frames in which faces are blurred or not sufficiently frontal-facing. (b) We discard segments with noisy speech by estimating speech SNR (see Section~\ref{sec:dataset}). The plot is intended to show the accuracy of our speech SNR estimator (and thus the quality of the dataset). We compare true speech SNR with our predicted SNR for synthetic mixtures of clean speech and non-speech noise at known SNR levels. Predicted SNR values (in dB) are averaged over $60$ generated mixtures per SNR bin, with error bars representing 1 std. We discard segments for which the predicted speech SNR is below 17 dB (marked by the gray dotted line in the plot).}
%
%
\label{fig:dataset_pipeline}
\afterfigure
\end{figure}

\paragraph{Dataset creation pipeline.}
%
%
Our dataset collection process has two main stages, as illustrated in Figure \ref{fig:dataset_pipeline}. First, we used the speaker tracking method of \citet{Hoover2017PuttingAF} to detect video segments of a person actively speaking with their face visible. Face frames that were blurred, insufficiently illuminated or had extreme pose were discarded from the segments. If more than 15\% of a segment's face frames were missing, it was discarded altogether. We used Google Cloud Vision API\footnote{https://cloud.google.com/vision/} for the classifiers in this stage, and to compute the statistics in Figure \ref{fig:dataset}.

The second step in building the dataset is refining the speech segments to include only clean, non-interfered speech. This is a crucial component because such segments serve as ground truth during training. We perform this refinement step automatically by estimating the speech SNR (the log ratio of the main speech signal to the rest of the audio signal) of each segment as follows.

We used a pre-trained audio-only speech denoising network to predict the SNR of a given segment using the denoised output as an estimation of the clean signal. \markedit{The architecture of this network is the same as the one implemented for the audio-only speech enhancement baseline in Section \ref{sec:exp}, and it was trained on speech from the LibriVox collection of public domain audio books.} Segments for which the estimated SNR is below a threshold were rejected. The threshold was set empirically using synthetic mixtures of clean speech and non-speech interfering noise at different, known SNR levels.\footnote{Such mixtures simulate well the type of interference in our dataset, which typically involves a single speaker interfered by non-speech sounds like audience clapping or intro music.} These synthetic mixtures were fed into the denoising network and the estimated (denoised) SNR was compared to the ground truth SNR (see Figure.~\ref{fig:dataset_pipeline}(b)).

We found that at low SNRs, on average, the estimated SNR is very accurate,  thus can be considered a good predictor of the original noise level. At higher SNRs (i.e. segments with little-to-no interference of the original speech signal), the accuracy of this estimator diminishes because the noise signal is faint. The threshold at which this occurs is at around 17 dB, as can be seen in Figure \ref{fig:dataset_pipeline}(b).
We listened to a random sample of 100 clips which passed this filtering, and found that none of them contained noticeable background noise. We provide sample video clips from our dataset in the supplementary material.

\section{Audio-Visual Speech Separation Model}
\label{sec:model}
\begin{figure*}
   \centering
   \includegraphics[width=.96\linewidth]{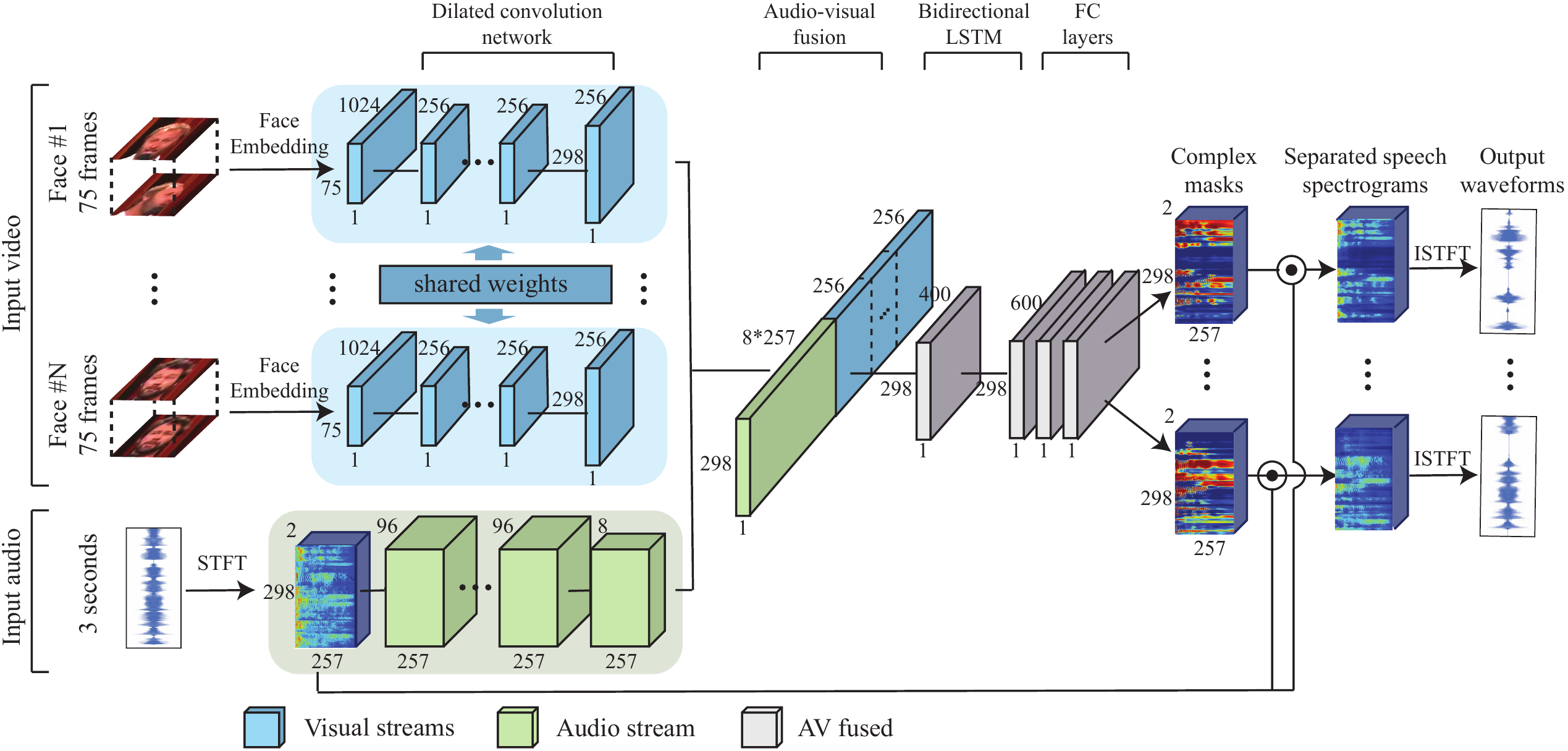}
   \vspace{-.1in}
   \caption{{\bf Our model's multi-stream neural network-based architecture:} The visual streams take as input thumbnails of detected faces in each frame in the video, and the audio stream takes as input the video's soundtrack, containing a mixture of speech and background noise. The visual streams extract face embeddings for each thumbnail using a pretrained face recognition model, then learn a visual feature using a dilated convolutional NN. The audio stream first computes the STFT of the input signal to obtain a spectrogram, and then learns an audio representation using a similar dilated convolutional NN. A joint, audio-visual representation is then created by concatenating the learned visual and audio features, and is subsequently further processed using a bidirectional LSTM and three fully connected layers. The network outputs a complex spectrogram mask for each speaker, which is multiplied by the noisy input, and converted back to waveforms to obtain an isolated speech signal for each speaker.} 
\label{fig:model}
\end{figure*}

At a high-level, our model is comprised of a multi-stream architecture which takes visual streams of detected faces and noisy audio as input, and outputs complex spectrogram masks, one for each detected face in the video (Figure~\ref{fig:model}). The noisy input spectrograms are then multiplied by the masks to obtain an isolated speech signal for each speaker, while suppressing all other interfering signals.

\begin{table*}
\centering
\caption{Dilated convolutional layers comprising our model's audio stream.}
\resizebox{\textwidth}{!}{%
\begin{tabular}{@{}lccccccccccccccc@{}}
& \tt conv1 & \tt conv2 & \tt conv3 & \tt conv4 & \tt conv5 & \tt conv6 & \tt conv7 & \tt conv8 & \tt conv9 & \tt conv10 & \tt conv11 & \tt conv12 & \tt conv13 & \tt conv14 & \tt conv15\\
\midrule
Num Filters & 96 & 96 & 96 & 96 & 96 & 96 & 96 & 96 & 96 & 96 & 96 & 96 & 96 & 96 & 8\\
Filter Size & $1 \times 7$ & $7 \times 1$ & $5 \times 5$ & $5 \times 5$   &  $5 \times 5$ & $5 \times 5$ & $5 \times 5$ & $5 \times 5$ & $5 \times 5$ & $5 \times 5$   &  $5 \times 5$ & $5 \times 5$ & $5 \times 5$ & $5 \times 5$ & $1 \times 1$ \\
Dilation & $1 \times 1$ & $1 \times 1$ & $1 \times 1$ & $2 \times 1$ & $4 \times 1$ & $8 \times 1$ & $16 \times 1$ & $32 \times 1$ & $1 \times 1$ & $2 \times 2$ & $4 \times 4$ & $8 \times 8$ & $16 \times 16$ & $32 \times 32$ & $1 \times 1$\\
Context    	& $1 \times 7$ & $7 \times 7$ & $9 \times 9$ & $13 \times 11$ & $21 \times 13$ & $37 \times 15$ & $69 \times 17$ & $133 \times 19$ & $135 \times 21$ & $139 \times 25$ & $147 \times 33$ & $163 \times 49$ & $195 \times 81$ & $259 \times 145$ & $259 \times 145$\\
\end{tabular}}
\label{tb:audio_stream}
\vspace{-.1in}
\end{table*}

\begin{table}
\centering
\caption{Dilated convolutional layers comprising our model's visual streams.}
\resizebox{0.47\textwidth}{!}{%
\begin{tabular}{lcccccc}
& \tt conv1 & \tt conv2 & \tt conv3 & \tt conv4 & \tt conv5 & \tt conv6\\
\midrule
Num Filters & 256 & 256 & 256 & 256 & 256 & 256 \\
Filter Size & $7 \times 1$ & $5 \times 1$ & $5 \times 1$ & $5 \times 1$   &  $5 \times 1$ & $5 \times 1$ \\
Dilation & $1 \times 1$ & $1 \times 1$ & $2 \times 1$ & $4 \times 1$ & $8 \times 1$ & $16 \times 1$ \\
Context    	& $7 \times 1$ & $9 \times 1$ & $13 \times 1$ & $21 \times 1$ & $37 \times 1$ & $69 \times 1$\\
\end{tabular}}
\label{tb:visual_stream}
\vspace{-.2in}
\end{table}

\subsection{Video and Audio Representation}
\label{ssec:data}

\paragraph{Input features.} \label{sec:features}
Our model takes both visual and auditory features as input. \markedit{Given a video clip containing multiple speakers, we use an off-the-shelf face detector (e.g. Google Cloud Vision API) to find faces in each frame (75 face thumbnails altogether per speaker, assuming 3-second clips at 25 FPS). We use a pretrained face recognition model to extract one face embedding per frame for each of the detected face thumbnails.} We use the lowest layer in the network that is not spatially varying, similar to the one used by \citet{cole2017synthesizing} for synthesizing faces. The rationale for this is that these embeddings retain information necessary for recognizing millions of faces, while discarding irrelevant variation between images, such as illumination. In fact, recent work also demonstrated that it is possible to recover facial expressions from such embeddings \cite{rudd2016moon}. We also experimented with raw pixels of the face images, which did not lead to improved performance.

As for the audio features, we compute the short-time Fourier transform (STFT) of 3-second audio segments. Each time-frequency (TF) bin contains the real and imaginary parts of a complex number, both of which we use as input. We perform power-law compression to prevent loud audio from overwhelming soft audio. The same processing is applied to both the noisy signal and the clean reference signal.

At inference time, our separation model can be applied to arbitrarily long segments of video. When more than one speaking face is detected in a frame, our model can accept multiple face streams as input, as we will discuss shortly.

\paragraph{Output.}
The output of our model is a multiplicative spectrogram mask, which describes the time-frequency relationships of clean speech to background interference. In previous work \cite{wang2014training,ss_overview}, multiplicative masks have been observed to work better than alternatives such as direct prediction of spectrogram magnitudes or direct prediction of time-domain waveforms. Many types of masking-based training targets exist in the source separation literature \cite{ss_overview}, of which we experiment with two: ratio mask (RM) and complex ratio mask (cRM).

The ideal ratio mask is defined as the ratio between the magnitudes of the clean and noisy spectrograms, and is assumed to lie between 0 and $1$.
The complex ideal ratio mask is defined as the ratio of the complex clean and noisy spectrograms. The cRM has a real component and an
imaginary component, which are separately estimated in the real domain. Real and imaginary parts of the complex mask will typically lie between -1 and 1, however, we use sigmoidal compression to bound these complex mask values between 0 and 1 \cite{Wang2016OraclePI}.

When masking with cRM, denoised waveforms are obtained by performing inverse STFT (ISTFT) on the complex multiplication of the predicted cRM and noisy spectrogram. When using RM, we perform ISTFT on the point-wise multiplication of the predicted RM and noisy spectrogram magnitude, combined with the noisy original phase \cite{ss_overview}.

Given multiple detected speakers' face streams as input, the network outputs a separate mask for each speaker, and one for background interference. We perform most of our experiments using cRM, as we found that output speech quality using it was significantly better than RM. See Table \ref{tb:ablation} for a quantitative comparison of the two methods.

\subsection{Network architecture}
\label{ssec:arch}
Fig.~\ref{fig:model} provides a high-level overview of the various modules in our network, which we will now describe in detail.

\paragraph{Audio and visual streams.}
The audio stream part of our model consists of dilated convolutional layers, the parameters of which are specified in Table~\ref{tb:audio_stream}.

The visual stream of our model is used to process the input face embeddings (see Section~\ref{ssec:data}), and consists of dilated convolutions as detailed in Table \ref{tb:visual_stream}. Note that ``spatial'' convolutions and dilations in the visual stream are performed over the temporal axis (not over the 1024-D face embedding channel).

To compensate for the sampling rate discrepancy between the audio and video signals, we upsample the output of the visual stream to match the spectrogram sampling rate ($100$ Hz). This is done using simple nearest neighbor interpolation in the temporal dimension of each visual feature.


\paragraph{AV fusion.}
The audio and visual streams are combined by concatenating the feature maps of each stream, which are subsequently fed into a BLSTM followed by three FC layers. The final output consists of a complex mask (two-channels, real and imaginary) for each of the input speakers. The corresponding spectrograms are computed by complex multiplication of the noisy input spectrogram and the output masks.  The squared error (L2) between the \markedit{power-law compressed} clean spectrogram and the enhanced spectrogram is used as a loss function to train the network. The final output waveforms are obtained using ISTFT, as described in Section~\ref{ssec:data}.

\paragraph{Multiple speakers.}
Our model supports isolation of multiple visible speakers in a video, each represented by a visual stream, as illustrated in Fig.~\ref{fig:model}. A separate, dedicated model is trained for each number of visible speakers, e.g. a model with one visual stream for one visible speaker, double visual stream model for two, etc. All the visual streams share the same weights across convolutional layers. In this case, the learned features from each visual stream are concatenated with the learned audio features before continuing on to the BLSTM. It should be noted that in practice, a model which takes a single visual stream as input can be used in the general case in which either the number of speakers is unknown, or a dedicated multi-speaker model is unavailable.

\subsection{Implementation details}
\label{ssec:details}

Our network is implemented in TensorFlow, and its included operations are used for performing waveform and STFT transformations. ReLU activations follow all network layers except for last (mask), where a sigmoid is applied. Batch normalization \cite{Ioffe2015BatchNA} is performed after all convolutional layers. Dropout is not used, as we train on a large amount of data and do not suffer from overfitting. We use a batch size of 6 samples and train with Adam optimizer for 5 million steps (batches) with a learning rate of $3\cdot10^{-5}$ which is reduced by half every 1.8 million steps.

All audio is resampled to 16kHz, and stereo audio is converted to mono by taking only the left channel. STFT is computed using a Hann window of length 25ms, hop length of 10ms, and FFT size of 512, resulting in an input audio feature of $257 \times 298 \times 2$ scalars. Power-law compression is performed with $p=0.3$ ($A^{0.3}$, where $A$ is the input/output audio spectrogram).

We resample the face embeddings from all videos to 25 frames-per-second (FPS) before training and inference by either removing or replicating embeddings. This results in an input visual stream of 75 face embeddings. Face detection, alignment and quality assessment is performed using the tools described by \citet{cole2017synthesizing}. When missing frames are encountered in a particular sample, we use a vector of zeros in lieu of a face embedding.

\section{Experiments and Results} \label{sec:exp}

We tested our method in a variety of conditions and also compared our results to state-of-the-art audio-only (AO) and audio-visual (AV) speech separation and enhancement, both quantitatively and qualitatively.

\paragraph{Comparison with Audio-Only}
There are no publicly available state-of-the-art audio-only speech enhancement/separation systems, and relatively few publicly available datasets for training and evaluating audio-only speech enhancement.  And although there is extensive literature on ``blind source separation'' for audio-only speech enhancement and separation \cite{comon2010handbook}, most of these techniques require multiple audio channels (multiple microphones), and are therefore not applicable to our task.  For these reasons, we implemented an AO baseline for speech enhancement which has a similar architecture to the audio stream in our audio-visual model (Fig.~\ref{fig:model}, when stripping out the visual streams).  When trained and evaluated on the CHiME-2 dataset \cite{Vincent2013TheS}, which is widely used for speech enhancement work, our AO baseline achieved a signal-to-distortion ratio of 14.6 dB, nearly as good as the state-of-the-art single channel result of 14.75  dB reported by \citet{Erdogan2015PhasesensitiveAR}.  Our AO enhancement model is therefore deemed a near state-of-the-art baseline.

In order to compare our separation results to those of a state-of-the-art AO model, we implemented the \textit{permutation-invariant training} introduced by~\citet{Yu2017PermutationIT}. Note that speech separation using this method requires \textit{a priori} knowledge of the number of sources present in the recording, and also requires \emph{manual} assignment of each output channel to the face of its corresponding speaker in the video (which our AV method does automatically).

We use these AO methods in all our synthetic experiments in Section~\ref{sec:synthetic}, and also show qualitative comparisons to it on real videos in Section~\ref{sec:results}. 

\paragraph{Comparison with Recent Audio-Visual Methods}
Since existing AV speech separation and enhancement methods are speaker dependent, we could not easily compare to them in our experiments on synthetic mixtures (Section~\ref{sec:synthetic}), or run them on our natural videos (Section~\ref{sec:results}). However, we show quantitative comparisons with those methods on existing datasets by running our model on videos from those papers. We discuss this comparison in more detail in Section~\ref{sec:comparison_with_av}. In addition, we show qualitative comparisons in our supplementary material.

\begin{table}
\small
\centering
\caption{{\bf Quantitative analysis and comparison with audio-only speech separation and enhancement:} Quality improvement (in SDR, see Section~\ref{sec:metrics} in the Appendix) as function of the number of input visual streams using different network configurations. First row (audio-only) is our implementation of a state-of-the-art speech separation model, and shown as a baseline.}
\begin{tabular}{lcccc}
\toprule[1.5pt]
& \bf 1S+Noise &\bf 2S clean &\bf 2S+Noise &\bf 3S clean \\
\midrule
AO \cite{Yu2017PermutationIT} & \bf 16.0 & 8.6 & 10.0 & 8.6 \\
\midrule
AV - 1 face  & \bf 16.0 & 9.9 & 10.1 & 9.1 \\
AV - 2 faces & - & \bf 10.3 & \bf10.6 & 9.1 \\
AV - 3 faces & - & - & - &  \bf10.0  \\
\bottomrule[1.5pt] \\
\end{tabular}
\vspace{-.15in}
\label{tb:main_comp}
\afterfigure
\end{table}

\subsection{Quantitative Analysis on Synthetic Mixtures} \label{sec:synthetic}

We generated data for several different single-channel speech separation tasks. Each task requires its own unique configuration of mixtures of speech and non-speech background noise. We describe below the generation procedure for each variant of training data, as well as the relevant models for each task, which were trained from scratch.

In all cases, clean speech clips and corresponding faces are taken from our \textsc{\datasetname} (AVS) dataset. Non-speech background noise is obtained from AudioSet \cite{audioset}, a large-scale dataset of manually-annotated segments from YouTube videos. Separated speech quality is evaluated using signal-to-distortion ratio (SDR) improvement from the BSS Eval toolbox \cite{BSSeval}, a commonly used metric for evaluating speech separation quality (see Section~\ref{sec:metrics} in the Appendix).

\markedit{We extracted 3-second non-overlapping segments from the varying-length segments in our dataset (e.g. a 10-sec segment would contribute 3 3-second segments). We generated 1.5 million synthetic mixtures for all the models and experiments. For each experiment, 90\% of the generated data was taken to be the training set, and the remaining 10\% was used as the test set.} We did not use any validation set as no parameter tuning or early stopping were performed.

\paragraph{One speaker + noise (1S+Noise).}
This is a classic speech enhancement task, for which the training data was generated by a linear combination of \markedit{unnormalized} clean speech and AudioSet noise: \markedit{$Mix_{i}=AVS_{j}+0.3*AudioSet_{k}$ where $AVS_{j}$ is one utterance from AVS, $AudioSet_{k}$ is one segment from AudioSet with its amplitude multiplied by 0.3, and $Mix_{i}$ is a sample in the generated dataset of synthetic mixtures.} Our audio only model performs quite well in this case, because the characteristic frequencies of noise are typically well separated from the characteristic frequencies of speech. Our audio-visual (AV) model performs as well as the audio-only (AO) baseline with SDR of 16 dB (first column of Table~\ref{tb:main_comp}).

\paragraph{Two clean speakers (2S clean).}
The dataset for this two-speaker separation scenario was generated by mixing clean speech of two different speakers from our AVS dataset: \markedit{$Mix_{i}=AVS_{j}+AVS_{k}$, where $AVS_{j}$ and $AVS_{k}$ are clean speech samples from different source videos in our dataset, and $Mix_{i}$ is a sample in the generated dataset of synthetic mixtures.} We trained two different AV models on this task, in addition to our AO baseline:

($i$) A model which takes only one visual stream as input, and outputs only its corresponding denoised signal. In this case, at inference, the denoised signal of each speaker is obtained by two forward passes in the network (one for each speaker). Averaging the SDR results of this model gives an improvement of 1.3 dB over our AO baseline (second column of Table~\ref{tb:main_comp}).

($ii$) A model which takes visual information from both speakers as input, in two separate streams (as explained in Section~\ref{sec:model}). In this case, the output consists of two masks, one for each speaker, and inference is done with a single forward pass. An additional boost of 0.4 dB is obtained using this model, resulting in a 10.3 dB total SDR improvement. Intuitively, jointly processing two visual streams provides the network with more information and imposes more constraints on the separation task, hence improving the results.

\markedit{Fig.~\ref{fig:sdr_scatter} shows the SDR improvement as a function of input SDR for this task, for both the audio-only baseline and our two-speaker audio-visual model.

\begin{figure}[t]
\centering
   \includegraphics[width=.8\columnwidth]{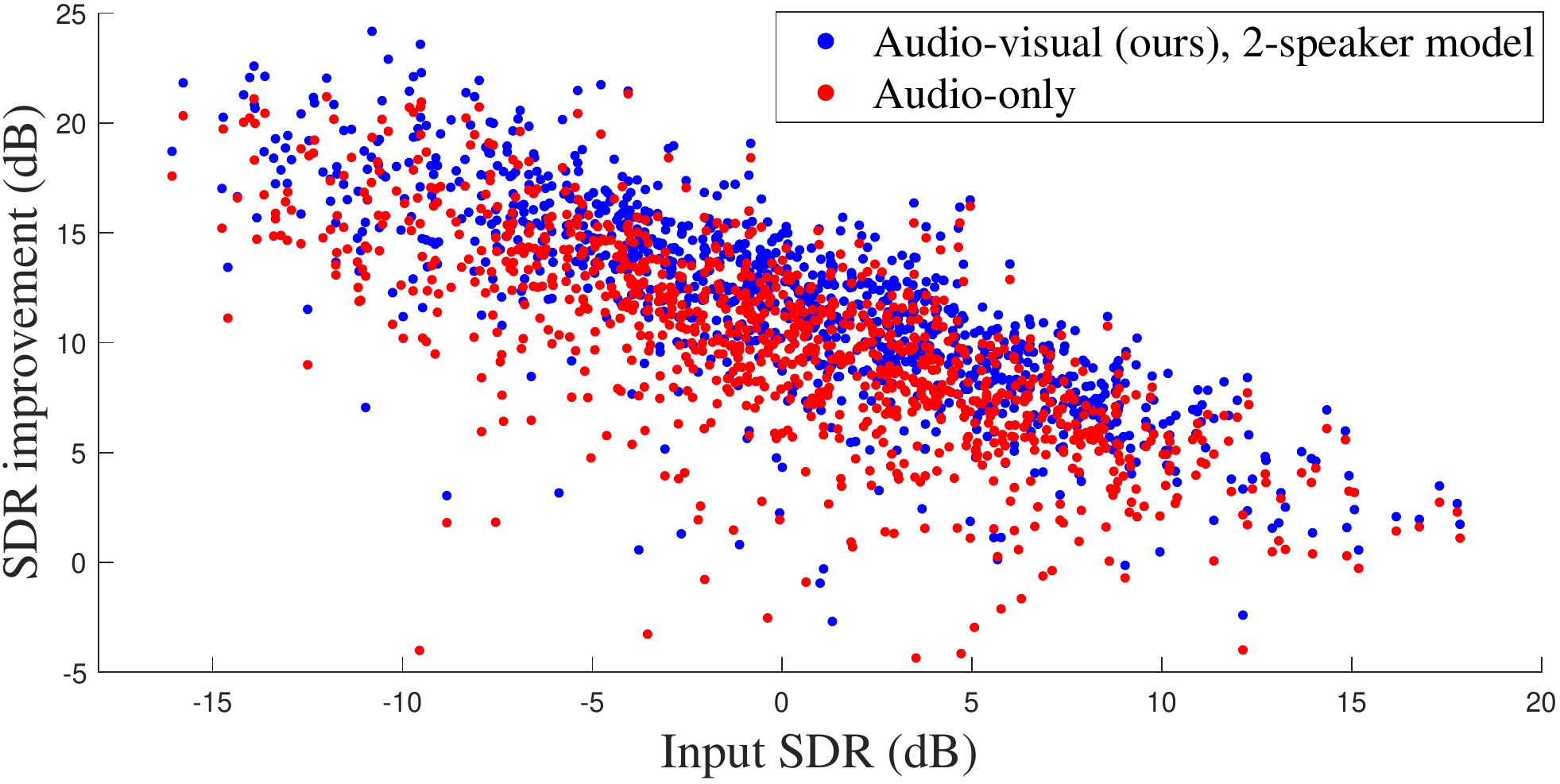}
\caption{\markedit{{\bf Input SDR vs. output SDR improvement:} A scatter plot showing separation performance (SDR improvement) as a function of original (noisy) SDR for the task of separating two clean speakers (\emph{2S clean}). Each point corresponds to a single, 3-second audio-visual sample from the test set.}}
\label{fig:sdr_scatter}
\afterfigure
\end{figure}
}

\begin{figure}
\centering
   \includegraphics[width=\columnwidth]{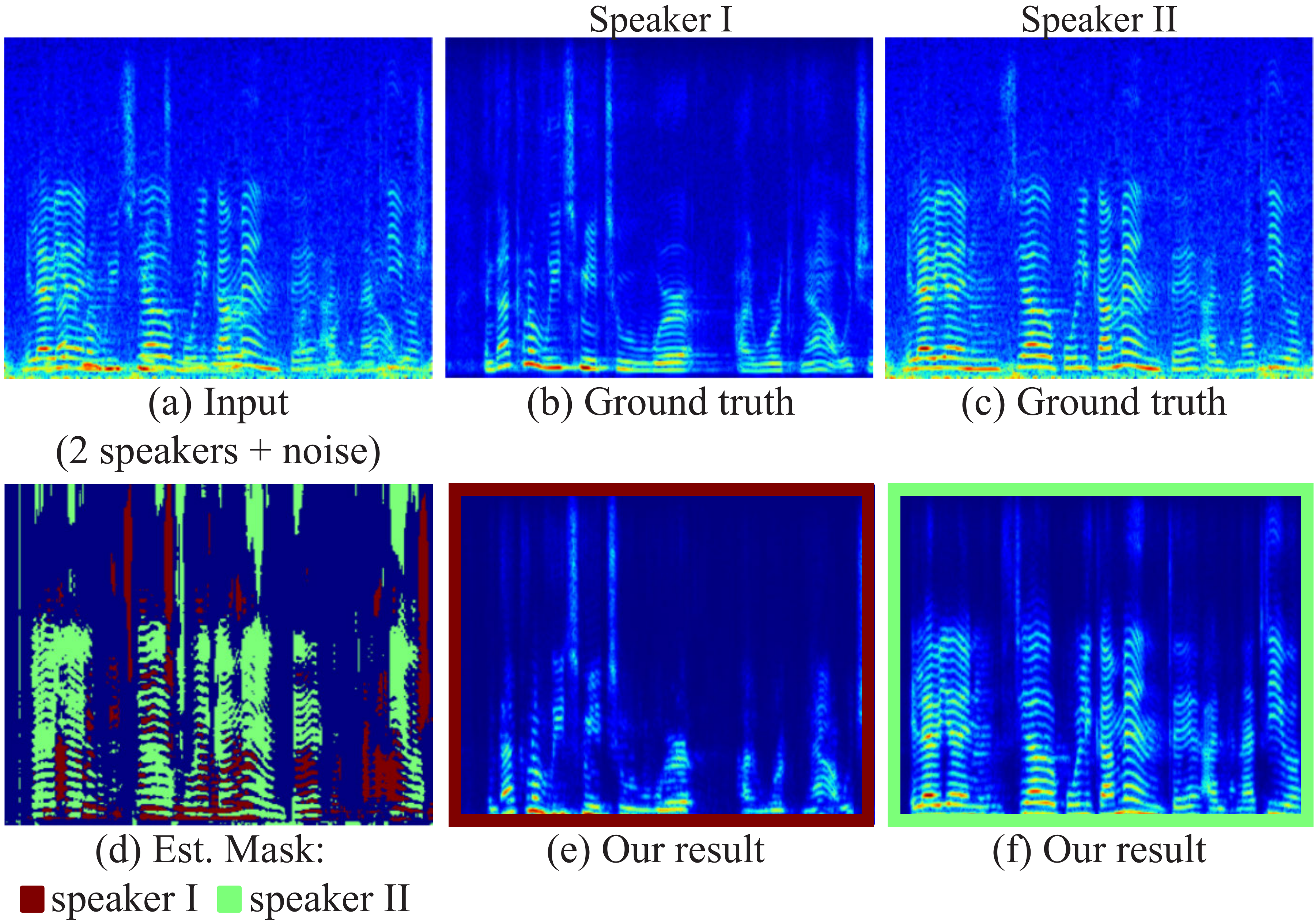}
%
%
\caption{{\bf Example of input and output audio:} The top row shows the audio spectrogram for one segment in our training data, involving two speakers and background noise (a), together with the ground truth, separate spectrograms of each speaker (b, c). In the bottom row we show our results: the masks our method estimates for that segment, superimposed on one spectrogram with a different color for each speaker (d), and the corresponding output spectrograms for each speaker (e, f).
}
\label{fig:specgrams}
\afterfigure
\end{figure}

\paragraph{Two speakers + noise (2S+Noise).}
Here, we consider the task of isolating one speaker's voice from a mixture of two speakers and non-speech background noise. To the best of our knowledge, this audio-visual task has not been addressed before. The training data was generated by mixing clean speech of two different speakers (as generated for the \emph{2S clean} task) with background noise from AudioSet: \markedit{$Mix_{i}=AVS_{j}+AVS_{k}+0.3*AudioSet_{l}$.}

In this case we trained the AO network with three outputs, one for each speaker and one for background noise. In addition, we trained two different configurations of our model, with one and two visual streams received as input. The configuration of the one-stream AV model is the same as in model ($i$) in the previous experiment. The two-stream AV outputs three signals, one for each speaker and one for background noise. As can be seen in Table~\ref{tb:main_comp} (third column), the SDR gain of our one-stream AV model over the audio only baseline is 0.1 dB, and 0.5 dB for two streams, bringing the total SDR improvement to 10.6 dB. Fig.~\ref{fig:specgrams} shows the inferred masks and output spectrograms for a sample segment from this task, along with its noisy input and ground truth spectrograms.

\paragraph{Three clean speakers (3S clean).}
The dataset for this task is created by mixing clean speech from three different speakers: \markedit{$Mix_{i}=AVS_{j}+AVS_{k}+AVS_{l}$.} In a similar manner to the previous tasks, we trained our AV model with one, two and three visual streams as input, which output one, two and three signals, respectively.

We found that even when using a single visual stream, the AV model performs better than the AO model, with a 0.5 dB improvement over it. The two visual stream configuration gives the same improvement over the AO model, while using three visual streams leads to a gain of 1.4 dB, attaining a total 10 dB SDR improvement (fourth column of Table~\ref{tb:main_comp}).

\paragraph{Same-gender separation.}
Many previous speech separation methods show a drop in performance when attempting to separate speech mixtures containing same-gender speech \cite{Hershey2016DeepCD,Delfarah2017FeaturesFM}. Table \ref{tb:gender_comp} shows a breakdown of our separation quality by the different gender combinations. Interestingly, our model performs best (by a small margin) on female-female mixtures, but performs well on the other combinations as well, demonstrating its gender robustness.

\begin{table}
  \begin{minipage}[c]{0.5\columnwidth}
    \caption{\textbf{Same-gender separation.} The results in this table, from the \emph{2S clean} experiment, show that our method is robust to separation of speech from same-gender mixtures.} \label{tb:gender_comp}
  \end{minipage} \hspace{.1in}
  \begin{minipage}[c]{0.4\columnwidth}
  	\small
    \centering
    \begin{tabular}{lc}
	\toprule[1.5pt]
	& \bf SDR \\
	\midrule
	Male-Male		&  9.7     \\
	Female-Female 	&  10.6     \\
	Male-Female	   	&  10.5  \\
	\bottomrule[1.5pt] \\
    \end{tabular}
  \end{minipage}
\vspace{-.3in}
\end{table}

\subsection{Real-World Speech Separation} \label{sec:results}
\begin{figure}
\centering
   \includegraphics[width=\columnwidth]{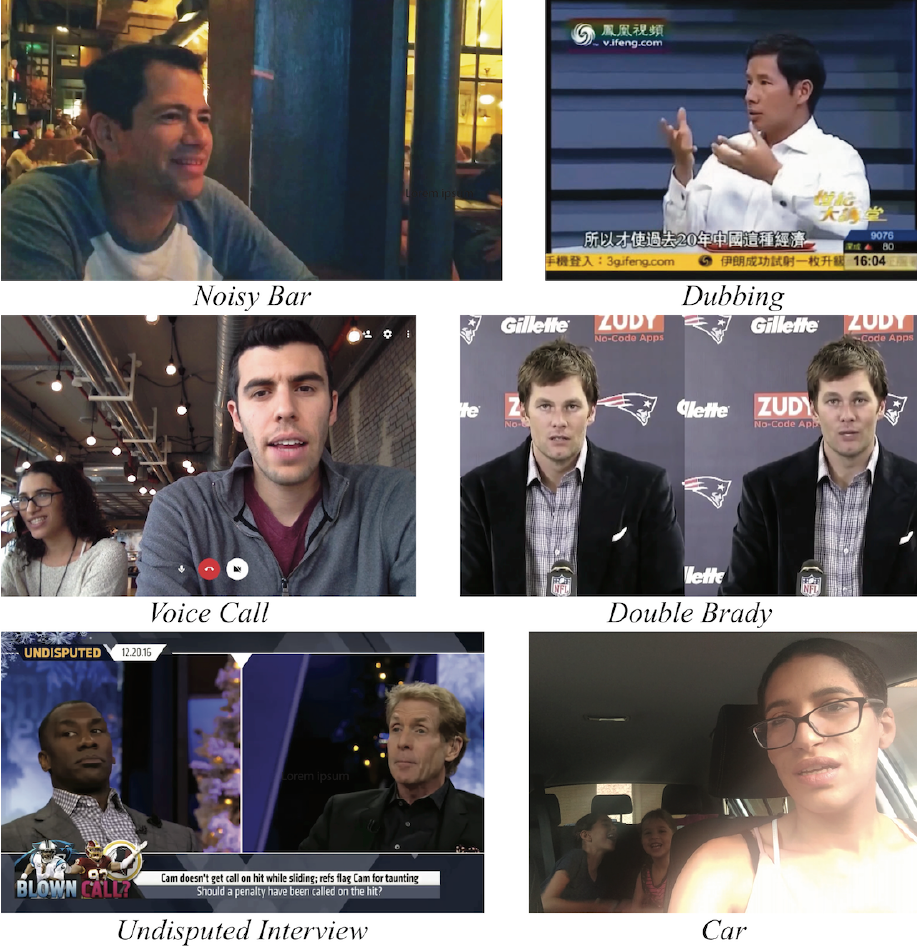}
   \vspace{-.2in}
\caption{{\bf Speech separation in the wild:} Representative frames from natural videos demonstrating our method in various real-world scenarios. All videos and results can be found in the supplementary material. The \emph{``Undisputed Interview''} video is courtesy of Fox Sports.}
\label{fig:sequences}
\afterfigure
\end{figure}

%
%
%
%
In order to demonstrate our model's speech separation capabilities in real-world scenarios, we tested it on an assortment of videos containing heated debates and interviews, noisy bars and screaming children (Fig.~\ref{fig:sequences}). In each scenario we use a trained model whose number of visual input streams matches the number of visible speakers in the video. \markedit{For example, for a video with two visible speakers, a two-speaker model was used. We performed separation using a single forward pass per video, which our model supports, since our network architecture never enforces a specific temporal duration.} This allows us to avoid the need to post-process and consolidate results on shorter chunks of the video. Because there is no clean reference audio for these examples, these results and their comparison to other methods are evaluated qualitatively; they are presented in our supplementary material. \markedit{It should be noted that our method does not work in real-time, and, in its current form, our speech enhancement is better suited for the post-processing stage of video editing.}

The synthetic \emph{``Double Brady''} video in our supplementary material highlights the utilization of visual information by our model, as it is very difficult to perform speech separation in this scenario using only characteristic speech frequencies contained in the audio. 

The \emph{``Noisy Bar''} scene shows a limitation of our approach in separating speech from mixtures with low SNR. In this case, the background noise is almost entirely suppressed, however output speech quality is noticeably degraded. \citet{Sun2017MultipletargetDL} observed that this limitation stems from the use of a masking-based approach for separation, and that in this scenario, directly predicting the denoised spectrogram could help overcome this problem. In cases of classic speech enhancement, i.e. one speaker with non-speech background noise, our AV model obtains similar results to those of our strong AO baseline. We suspect this is because the characteristic frequencies of noise are typically well separated from the characteristic frequencies of speech, and therefore incorporating visual information does not provide additional discrimination capabilities.

\subsection{Comparison with Previous Work in Audio-Visual Speech Separation and Enhancement}
\label{sec:comparison_with_av}
Our evaluation would not be complete without comparing our results to those of previous work in AV speech separation and enhancement. Table \ref{tb:method_comp} contains these comparisons on three different AV datasets, Mandarin, TCD-TIMIT and CUAVE, mentioned in Section~\ref{sec:related}, using the evaluation protocols and metrics described in the respective papers. The reported objective quality scores are PESQ \cite{rix2001perceptual}, STOI \cite{taal2010short} and SDR from the BSS eval toolbox \cite{BSSeval}. Qualitative results of these comparisons are available on our project page.

It is important to note that these prior methods require training a dedicated model for each speaker in their dataset (\emph{speaker dependent}), whereas our evaluation on their data is done using a model trained on our general AVS dataset (\emph{speaker independent}). Despite having never encountered these particular speakers before, our results are significantly better than those reported in  the original papers, indicating the strong generalization capability of our model.

\begin{table}
\small
\centering
\caption{\textbf{Comparison with existing audio-visual speech separation work.} We compare our speech separation and enhancement results on several datasets to those of previous work, using the evaluation protocols and objective scores reported in the original papers. Note that previous approaches are \emph{speaker-dependent}, whereas our results are obtained by using a general, \emph{speaker-independent} model.}
\begin{tabular}{lcccc }
\toprule[1.5pt]
\multicolumn{4}{c}{\bf Mandarin (Enhancement)} \\
\midrule
 &  \citet{gabbay2017visual} & \citet{Hou2017AudioVisualSE} & Ours \\
\midrule
PESQ & 2.25 & 2.42 & \bf{2.5} \\ 
STOI &  -   & 0.66 & \bf{0.71} \\
\markedit{SDR}  &  -   & 2.8  & \bf{6.1} \\
\midrule
\midrule
\multicolumn{4}{c}{\bf TCD-TIMIT (Separation)} \\
\midrule
 &  \citet{gabbay2017visual} & Ours \\
\midrule
\markedit{SDR} &  0.4 & \bf{4.1} &  \\
 PESQ & 2.03 & \bf{2.42} &  \\
\midrule
\midrule
\multicolumn{4}{c}{\bf CUAVE (Separation)} \\
\midrule
& \citet{casanovas2010blind} & \citet{pu2017audio} &  Ours \\
\midrule
\markedit{SDR} &  7 & 6.2 & \bf{12.6} \\ 
\bottomrule[1.5pt] \\
\end{tabular}
\vspace{-.15in}
\label{tb:method_comp}
\afterfigure
\end{table}

\subsection{Application to Video Transcription}
While our focus in this paper is speech separation and enhancement, our method can also be useful for automatic speech recognition (ASR) and video transcription. As a proof of concept, we performed the following qualitative experiment. We uploaded our speech-separated results for the \emph{``Stand-Up''} video to YouTube, and compared the resulting captions produced by YouTube's automatic captioning\footnote{\url{https://support.google.com/youtube/answer/6373554?hl=en}} with those it produced for the corresponding source videos with mixed speech. For parts of the original \emph{``Stand-Up''} video, the ASR system was unable to generate any captions in mixed speech segments of the video. The results included speech from both speakers, resulting in hard-to-read sentences. However, captions produced on our separated speech results were noticeably more accurate. We show the full captioned videos in our supplementary material.

\subsection{Additional Analysis} \label{ssec:anaysis}

We also conducted extensive experiments to better understand the model's behavior and how its different components affect the results.

\begin{table}[]
\small
\centering
\caption{\textbf{Ablation study:} We investigate the contribution of different parts of our model on the scenario of separating mixtures of two clean speakers. SDR correlates well with noise suppression, and ViSQOL indicates level of speech quality (see Section~\ref{sec:metrics} in the Appendix)}.
\begin{tabular}{lcc}
\toprule[1.5pt]
& \bf SDR & \bf ViSQOL \\
\midrule
Our full model (cRM) &  \bf 10.3 & \bf 3.0  \\
- No FC              &  10.0 & 2.3 \\
- No BLSTM           &  8.7 &  2.7 \\
- Audio-only (input) &   8.6  & 2.7 \\
- No BLSTM or FC 	 &  0.6  & 1.8  \\
- Visual-only (input)&  1.4  & 2.0 \\
\midrule
Magnitude mask (RM) &  9.7  & 2.8 \\
Bottleneck (cRM) 	&  9.8  & 2.9 \\
Early fusion (cRM)	&  8.1  & 2.5 \\
\midrule
Oracle RM + noisy ph. & 9.7 & 3.4 \\
Oracle cRM    &  14.8 &  3.8 \\
Oracle RM + oracle ph.  & 17.4  & 4.7  \\
\bottomrule[1.5pt] \\
\end{tabular}
\label{tb:ablation}
\vspace{-.3in}
\end{table}

\paragraph{Ablation study.}
In order to better understand the contribution of different parts of our model, we performed an ablation study on the task of speech separation from a mixture of two clean speakers (\emph{2S Clean}). In addition to ablating several combinations of network modules (visual and audio streams, BLSTM and FC layers), we also investigated higher-level changes such as a different output mask (magnitude), the effect of reducing the learned visual features to one scalar per timestep, and a different fusion method (early fusion).

In the early fusion model, we do not have separate visual and audio streams, but rather combine the two modalities at the input. This is done by first using two fully connected layers to reduce the dimensionality of each visual embedding to match the spectrogram dimension at each timestep, then stacking the visual features as a third spectrogram ``channel'' and processing them jointly throughout the model.
%
%
%
%
%
%

Table \ref{tb:ablation} shows the results of our ablation study. The table includes evaluation using SDR and ViSQOL \cite{Hines2015ViSQOLAudioAO}, an objective measure intended to approximate human listener mean opinion scores (MOS) of speech quality. The ViSQOL scores were calculated on a random 2000 sample subset of our testing data. We found that SDR correlates well with the amount of noise left in the separated audio, and ViSQOL is a better indicator of output speech quality. \markedit{See Section~\ref{sec:metrics} in the Appendix for more details on these scores. ``Oracle'' RMs and cRMs are masks obtained as described in Section \ref{ssec:data}, by using the ground truth real-valued and complex-valued spectrograms, respectively.}

The most interesting findings of this study are the drop in MOS when using a real-valued magnitude mask rather than a complex one, and the surprising effectiveness of squeezing the visual information into one scalar per timestep, described below.

\paragraph{Bottleneck features.}
In our ablation analysis we found that a network which squeezes the visual information into a bottleneck of one scalar per timestep (``Bottleneck (cRM)'') performs almost as well (only 0.5 dB less) as our full model (``Full model (cRM)'') that uses 64 scalars per timestep. 
%
%
%
%

\paragraph{How does the model utilize the visual signal?}
Our model uses face embeddings as the input visual representation (Section~\ref{sec:features}). We want to gain insights on the information captured in these high-level features and to identify which regions of the input frames are used by the model for separating the speech. To this end, we follow a similar protocol as in~\cite{zhou2014object, zeiler2014visualizing} for visualizing receptive fields of deep networks. We extend that protocol from 2D images to 3D (space-time) video. More specifically, we use a space-time patch occluder ($11px \times 11px \times 200ms$ patch\footnote{We use 200ms length to cover the typical range of phoneme duration: 30-200 ms.}) in a sliding window fashion. For each space-time occluder, we feed-forward the occluded video into our model and compare the speech separation result, $S_{occ}$, with the one obtained on the original (non-occluded) video, $S_{orig}$. To quantify the difference between the network outputs, we use SNR, treating the result without the occluder as the ``signal''\footnote{We refer the reader to the supplementary material to validate that our separated speech on the non-occluded video, which we treat as ``correct'' in this example, is indeed accurate.}. That is, for each space-time patch, we compute:
\begin{equation}
E=10 \cdot \log{\left(\frac{{S_{orig}}^2}{({S_{occ}-S_{orig}})^2}\right)}.
\end{equation}
Repeating this process for all space-time patches in a video results in a heat map for each frame. For visualization purposes we normalize the heat maps by the maximum SNR for the video: $\tilde{E} = E_{max} - E$.  In $\tilde{E}$, high values correspond to patches with high impact on the speech separation result.

\begin{figure}[t]
\centering
   \includegraphics[width=.92\columnwidth]{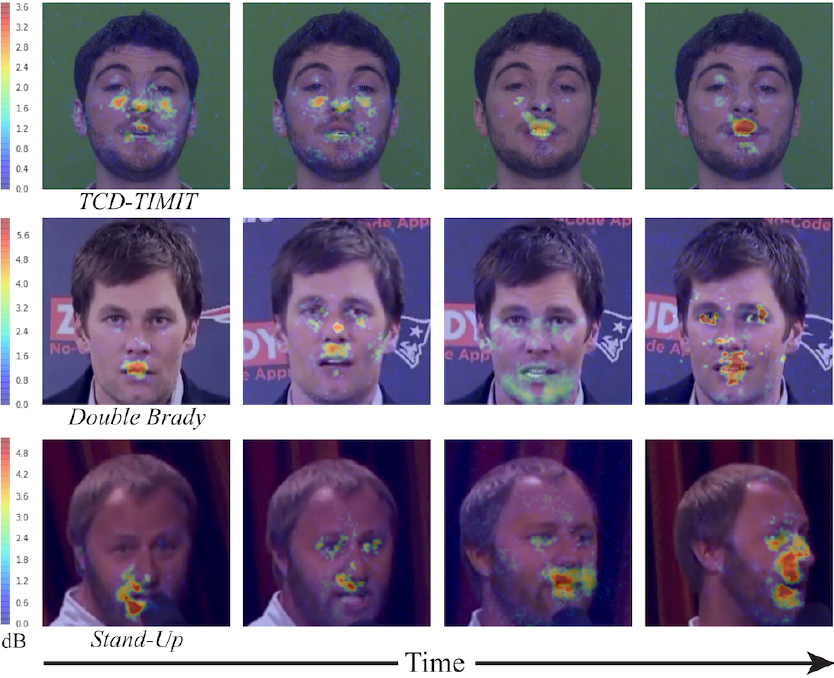}
\vspace{-.15in}
\caption{\markedit{{\bf How does the model utilize the visual signal?} We show heat maps overlaid on representative input frames from several videos, visualizing the contribution of different regions of the frames to our speech separation result (in dB, see text), from blue (low contribution) to red (high contribution).}}
\label{fig:visualization}
\afterfigure
\end{figure}

\markedit{In Fig.~\ref{fig:visualization} we show the resulting heat maps for representative frames from several videos (the full heat map videos are available on our project page). As expected, the facial regions that contribute the most are located around the mouth, yet the visualization reveals that other areas such as the eyes and cheeks contribute as well.}

\paragraph{Effect of missing visual information.}
We further tested the contribution of visual information to the model by gradual elimination of visual embeddings. Specifically, we start by running the model and evaluating the speech separation quality using visual information for the full 3 second video. \markedit{We then gradually discard embeddings from both ends of the segment, and re-evaluate the separation quality with visual durations of 2, 1, 0.5 and 0.2 seconds.}

\begin{figure}[t]
\centering
   \includegraphics[width=.7\columnwidth]{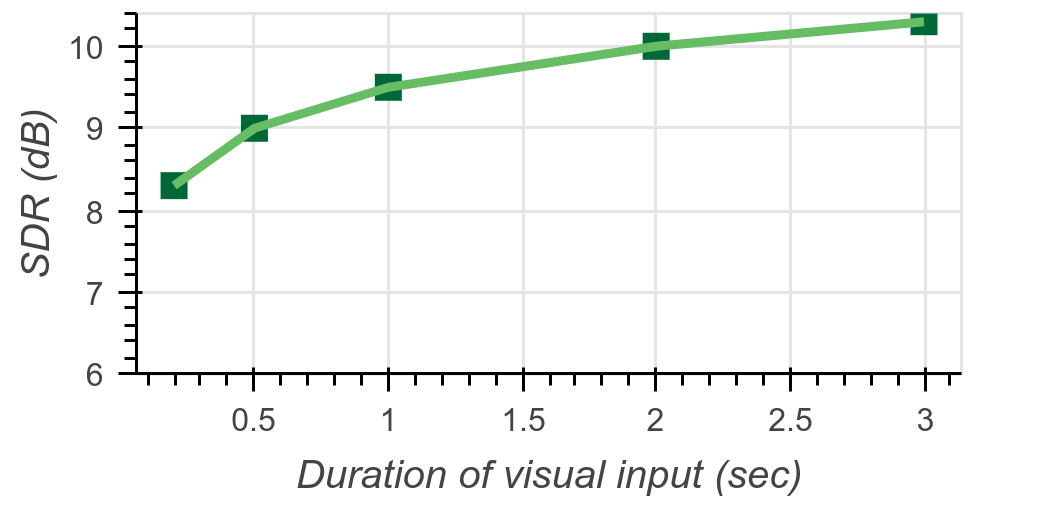}
\vspace{-.1in}
\caption{{\bf The effect of missing visual information:} This graph shows the impact of the duration of the visual information on output SDR improvement in the \emph{2S clean} scenario. We test this by gradually zeroing-out input face embeddings from both ends of the sample. The results show that even a small number of visual frames are sufficient for high-quality separation.}
\label{fig:elim}
\afterfigure
\end{figure}

The results are shown in Fig.~\ref{fig:elim}.
Interestingly, the speech separation quality is reduced by only $0.8$ dB on average when dropping as much as $2/3$ of the visual embeddings in the segments. This shows the robustness of the model to missing visual information, which may occur in real world scenarios due to head motion or occlusions.


\section{Conclusion} \label{sec:conc}

We proposed an audio-visual neural network-based model for single-channel, speaker-independent speech separation. Our model works well in challenging scenarios, including multi-speaker mixtures with background noise. To train the model, we created a new audio-visual dataset with thousands of hours of video segments containing visible speakers and clean speech we collected from the Web. We showed state-of-the-art results on speech separation as well as a potential application to video captioning and speech recognition. We also conducted extensive experiments to analyze the behavior of our model and its components.


\section*{Acknowledgements}
We would like to thank Yossi Matias and Google Research Israel for their support for the project, and John Hershey for his valuable feedback. We also thank Arkady Ziefman for his help with figure design and video editing, and Rachel Soh for helping us procure permissions for video content in our results.

\bibliographystyle{ACM-Reference-Format}
\bibliography{references}

\appendix
\markedit{
\section{Objective metrics used for evaluating separation quality}
\label{sec:metrics}

\subsection{SDR}
The signal-to-distortion ratio (SDR), introduced by \citet{BSSeval}, is one of a family of measures designed to evaluate Blind
Audio Source Separation (BASS) algorithms, where the original source signals are available as ground truth. The measures are based on the decomposition of each estimated source signal into a true source part ($s_{target}$) plus error terms corresponding to interferences ($e_{interf}$), additive noise ($e_{noise}$) and algorithmic
artifacts ($e_{artif}$).

SDR is the most general score, commonly reported for speech separation algorithms. It is measured in dB, and is defined as:
\begin{equation}
SDR:=10 \cdot \log_{10}{\left(\frac{||{s_{target}}||^2}{||{e_{interf}+e_{noise}+e_{artif}}||^2}\right)}.
\end{equation}
We refer the reader to the original paper for details on signal decomposition into its components. We found this measure to correlate well with the amount of noise left in the separated audio.

\subsection{ViSQOL}
The Virtual Speech Quality Objective Listener (ViSQOL) is an objective speech quality model, introduced by \citet{Hines2015ViSQOLAudioAO}. The metric models human speech quality perception using a spectro-temporal measure of similarity between a reference ($r$) and a degraded ($d$) speech signal, and is based on the Neurogram Similarity Index Measure (NSIM) \cite{nsim}. NSIM is defined as
\begin{equation}
NSIM(r,d)=\frac{2\mu_{r}\mu_{d}+C_{1}}{\mu_{r}^2+\mu_{d}^2+C_{1}}\cdot\frac{\sigma_{rd}+C_{2}}{\sigma_{r}\sigma_{d}+C_{2}},
\end{equation}
where the $\mu$s and $\sigma$s are mean and correlation coefficients, respectively, calculated between reference and degraded spectrograms.

In ViSQOL, NSIM is calculated on spectrogram patches of the reference signal and their corresponding patches from the degraded signal. The algorithm subsequently aggregates and translates the NSIM scores into a mean opinion score (MOS) between 1 and 5.

}

\end{document}